\documentclass{article}

\usepackage{lineno,hyperref}
\modulolinenumbers[5]

\usepackage{lineno,hyperref}
\modulolinenumbers[5]

\usepackage[english]{babel}
\usepackage[utf8x]{inputenc}
\usepackage{amsmath}
\usepackage{graphicx}
\usepackage[colorinlistoftodos]{todonotes}
\usepackage{amssymb}

\usepackage{gensymb}
\usepackage{newunicodechar}
\newunicodechar{°}{\degree}

\usepackage{url}

\usepackage{authblk}
\title{Addressing the eye-fixation problem in gaze tracking for human computer interface using the Vestibulo-ocular Reflex.}

\author{Adam~Pantanowitz}
\author[1]{Kimoon~Kim}
\author[1]{Chelsey~Chewins}
\author[1]{Isabel~N.~K.~Tollman}
\author[1]{David~M.~Rubin}

\affil[1]{School of Electrical and Information Engineering\\University of the Witwatersrand, Johannesburg\\1 Jan Smuts Avenue, Braamfontein, South Africa}

\begin{document}
\maketitle

\begin{abstract}

A custom head-mounted system to track smooth eye movements for control of a mouse cursor is implemented and evaluated. The system comprises a head-mounted infrared camera, an infrared light source, and a computer. Software-based image processing techniques, implemented in Microsoft Visual Studio, OpenCV, and Pupil, detect the pupil position and direction of pupil movement in near real-time. The identified direction is used to determine the desired positioning of the cursor, and the cursor moves towards the target. Two users participated in three tests to quantify the differences between incremental tracking of smooth eye movement resulting from the Vestibulo-ocular Reflex versus step-change tracking of saccadic eye movement. Tracking smooth eye movements was four times more accurate than tracking saccadic eye movements, with an average position resolution of 0.80~cm away from the target. In contrast, tracking saccadic eye movements was measured with an average position resolution of 3.21~cm. Using the incremental tracking of smooth eye movements, the user was able to place the cursor within a target as small as a 9~$\times$~9 pixel square~90~\% of the time. However, when using the step change tracking of saccadic eye movements, the user was unable to position the cursor within the 9~$\times$~9 pixel target. The average time for the incremental tracking of smooth eye movements to track a target was 6.45 s, whereas for the step change tracking of saccadic eye movements, it was 2.61~s.
\end{abstract}

\section{Introduction}
%
%
%
%
Traditional step change tracking of saccadic eye movements, in which the user’s gaze is determined using a camera mounted on a computer, exhibit inaccuracies for three reasons, as identified by~\cite{Hyrskykari_Utilizing_2006}, viz. a tolerance of approximately one degree in the point position, the accuracy of eye tracking being heavily dependent on the quality of the calibration performed, and emergent calibration drifts during the user’s session of controlling the cursor. Calibration drifts occur because the change in pupil size and head movements modify the validity of the calibration. In addition to these problems, tracking cursor motion while vision is fixated on a destination point is a step change with open-loop control. This presents a substantial challenge: the user’s gaze moves from the mouse cursor to the destination position, and does not track the position of the cursor as it moves.
Smooth eye movements resulting from the Vestibulo-ocular Reflex (VOR)~\cite{Tomlinson_Naso_2000} offers an alternative, slower, closed-loop control, where the user can observe the mouse cursor as it moves towards its destination. This closed-loop tracking occurs as the user is able to smoothly move the mouse cursor while monitoring its trajectory, until the cursor reaches its intended destination.
The presented system is capable of tracking VOR-related smooth eye movement for the purposes of comparison with saccadic eye movement tracking. This approach to exploiting the VOR for tracking smooth eye movements was first proposed by Jahnke~\cite{vorpatent}, wherein the camera is mounted to the user’s head, allowing the user to remain focused on the cursor’s position for tracking purposes, while maintaining the ability to achieve high deflection for large-scale eye movements.
Measurements were taken to compare incremental tracking of smooth eye movements and step change tracking of saccadic eye movements using a head-mounted system. This facilitates the development and evaluation of a custom eye tracking mouse cursor control system. The system’s control performance is measured and assessed to evaluate cursor tracking.



\section{Background}

Hutchinson et al. \cite{10_Hutchinson_Human_2017}  argue that the application of eye movements in user interfaces is a key area of research in Human Computer Interaction, as most people have control over eye movements regardless of disability. This can solve a variety of usability problems, such as: web usage; operation of mobile devices; software; and gaming systems  \cite{11_rosson2002usability}. Thus, improvements in controlling a mouse cursor through eye tracking can be expected to enhance eye tracking user interfaces.
\subsection{Eye Movements}
 Of the several eye movements described in various situations \cite{4_Wang_An_2013}, vergence movements are used to track an approaching object \cite{5_Zhang_A_2005}. When an object approaches, the eyes move medially. Faster saccadic movements are used to track fast-moving objects or when reading~\cite{6_Alkan_Functional_2011}. Similarly, rapid smooth pursuit movements are used to track slower moving objects. Saccade and smooth pursuit movements are used in traditional step change eye tracking algorithms. Smooth movements resulting from the VOR are used to focus on an object while the head moves in space (\cite{7_Quinn_Modeling_1990}). \cite{Tomlinson_Naso_2000} point out that smooth tracking produces slower eye movement than saccadic movements, as used, for example, when reading.
The VOR allows a user to stabilise an image on the retina’s field of view while the head moves in space (\cite{Tomlinson_Naso_2000}). There are two types of head movements for which systems in the brain compensate: rotations and translations. Rotational head movements are detected by the semi-circular canals in the inner ear, whereas translational head movements are determined by the otoliths in the inner ear.
The extra-ocular muscles serve as the actuators in the VOR which facilitate the stability of eye gaze (\cite{8_Goumans_Three_2010}). The image projected onto the retina is stabilised during head movement because the extra-ocular muscles rotate the eye in the opposite direction to head movement.

\subsection{Eye Tracking}
The first documented precise, non-invasive, eye tracking technique was performed by~\cite{12_Dodge_The_1901} using light reflected off the cornea. Since then, a variety of eye tracking techniques have been developed and tested. \\\\
Two such techniques, electro-oculography and video-oculography, have been used widely in research.
Electro-oculography is a technique for measuring the corneo-retinal action potential that exists due to the metabolic activity in the retina~\cite{13_malmivuo1995electric}. This creates a positive and negative pole at the cornea and retina, respectively.

Therefore, a small voltage difference exists between the front and back of the eyes which can be measured using electrodes placed in pairs either above and below the eye, or to the left and right of the eye. The measured voltage generally ranges from 0.4 to 1.0~mV. When the eyes move horizontally, the maximum achievable rotation is approximately $70~^{\circ}$, and thus creates a movement signal ranging from 5 to 20~$\mu V$. The signal is then amplified, filtered, and converted to digital signals to acquire usable data.
\cite{14_Journal_Video_2001} defines video-oculography as a technique to capture eye movement using a digital video camera. A variety of methods are used to capture eye movement in video-oculography, including the use of head-mounted gear or a pair of glasses equipped with a video camera(s) to capture eye movement. These techniques produce close-up images of the eye that are simple to process using image processing software.

A more challenging technique is to use a video camera mounted away from the head, such as a web camera mounted on a screen, to capture eye movements \cite{15_Chennamma_A_2013}. This requires significantly more image processing because the face and eyes must be detected before eye movement can be extracted~\cite{16_Chou_Real_2008}. Both techniques can be performed using either visible or infrared light~\cite{15_Chennamma_A_2013}, however, infrared light and a mounted camera may be preferable as the pupil shows up as a black circle on images and is easily detectable \cite{17_Punde_A_2017}.


\section{Methodology}

Exploiting slower, smooth movements for eye tracking provides the user with greater incremental control over the cursor and, unlike saccadic eye tracking, allows the user to observe the cursor while it is in motion. Closed-loop control is the primary benefit of smooth eye tracking. With saccadic eye tracking, the user must take their eyes off the cursor to move it. In contrast, smooth eye tracking determines the position of the eyes relative to the head, rather than the position on which the user’s gaze is fixated, while allowing for large scale deflection.
This has the potential to produce a number of benefits for the smooth cursor tracking control system: improved accuracy through closed-loop tracking; greater eye-movement comfort with less eye deflection; and greater robustness. Greater robustness occurs as tracking is a continuous incremental movement rather than erratic saccadic movement as exhibited in saccadic eye tracking.
The system must therefore allow the user to move a cursor using smooth eye tracking, and to provide full range of cursor movement. The hardware used must be compatible with standard and commonly used operating systems. To solve the challenges of saccadic eye tracking, the system must be capable of pupil detection and smooth eye movement tracking as the eyes move relative to the head in space.

\section{System Design}

\begin{figure}
\centering
\includegraphics[width=2.5in]{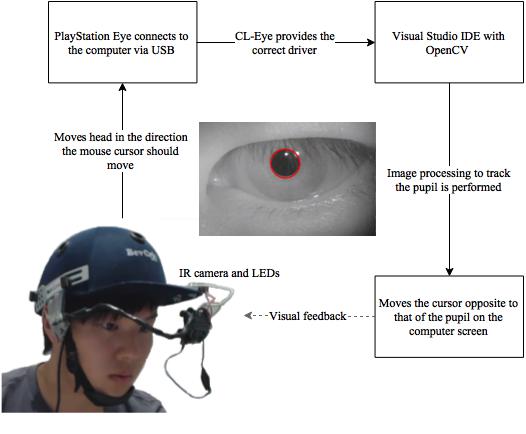}
\caption{System overview indicating all system elements}
\label{fig:SystemOverview}
\end{figure}

Image processing is constrained by the computer’s processing power. Electrical safety and set infrared expo-sure limits for eye safety are prioritised.
The system uses smooth head movements and eye tracking to move a mouse cursor across a computer screen. The system is divided into two subsystems: hardware and software. Figure~\ref{fig:SystemOverview} provides an overview of the entire system, including the wearable headgear, detection of pupil, hardware to software connection, and visual feedback.

\subsection{Hardware Subsystem}

The hardware subsystem consists of the PlayStation-Eye camera, headgear, and infrared circuit. Pupil movement is acquired using a modified PlayStation-Eye cam-era attached to headgear, as illustrated in Figure~\ref{fig:SystemOverview}. This creates wearable instrumentation which enables freedom of head movement. The camera is modified to receive exclusively infrared light. A separate circuit is built and attached to the headgear to provide an infrared light source. The PlayStation-Eye is connected to a computer via a Universal Serial Bus (USB). The Code Laboratories Eye (CL-Eye) platform driver, which is installed on the computer, provides a certified hardware driver so that third party software programs can access the PlayStation-Eye camera. Further information is available at~\cite{18_code_labs}.
The PlayStation-Eye camera has a built-in infrared light filter and lenses per the specification manufacturer, Sony Computer Entertainment~\cite{20_sony}. This is modified to capture close-up infrared images and videos.
According to Sony Computer Entertainment~\cite{20_sony}, the PlayStation-Eye can alter its zoom lens from a 56 ˚ field of view to a 75 ˚ field of view by using its built-in mechanism. However, using this information and basic trigonometry, depicted in Figure 2, the shortest field of view distance that can be achieved with a focal distance of 200 mm – for eye safety reasons – is 213 mm. This is large considering the size of the eye~\cite{bekerman2014variations}. Therefore, the PlayStation-Eye lens must be re-placed with a lens that accommodates a 60 mm rectangular region from the field of view at a focal distance of 200 mm. The chosen lens is a 60 mm m12 zoom lens that has a 23 ˚ field of view, as described by~\cite{21_peau}. This lens fulfils the necessary requirements and has an 81 mm field of view at a focal distance that will take adequate close-up pictures of the eye.



\begin{figure}[!t]
\centering
\includegraphics[width=2.5in]{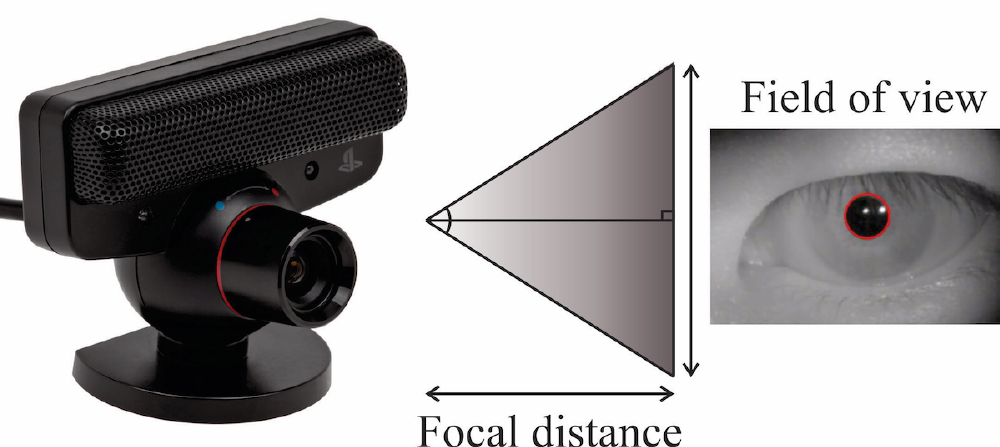}
\caption{Relationship between focal distance and field of view, adapted from~\cite{22_amos} and~\cite{23_natinstr}}
\label{fig:focaldistance}
\end{figure}

A visible light filter must be inserted between the image sensor and the lens so that only infrared light passes through. The visible light filter is constructed using magnetic tape cut out of a floppy disk~\cite{24_mcll}. This results in an infrared image of the eye shown in Figure 1.
The headgear is a cricket (sport) helmet, modified to provide structural support for the PlayStation-Eye camera and the infrared circuit. It ensures that the camera is directed at the eye as the head moves. The helmet’s grill visor is removed to eliminate obstructions to the eyes. A malleable copper wire is used to connect the PlayStation-Eye camera to the headgear. This ensures that the direction of the camera and the infrared light emitting diodes (LEDs) can be changed when different people use the headgear.
The infrared circuit provides infrared light to the eyes. It consists of an Atmel ATmega328p micro-controller connected to a 3 V source with two infrared LEDs connected to the micro-controller’s pulse width modulation (PWM) pins. The 3 V source is constructed by connecting two AA batteries, each of which has a nominal volt-age of 1.5 V, in series. The PWM duty cycle is set at 50 \% to further limit infrared exposure to the eyes. Electrical safety and eye safety exposure limits are prioritised in the design.


\subsection{Software Subsystem}

The software subsystem is responsible for moving the mouse cursor once movement of the pupil has been detected. This is implemented using a C++ integrated development environment (IDE), such as Microsoft Visual Studio 2013, with OpenCV 3.0 for image processing (an open source C++ library).

According to Code Laboratories (2014), the PlayStation-Eye captures video at a rate of 60 frames per second at 640~×~480 pixels. Therefore, the image processing algorithm is performed to detect the pupil position at each frame. The proposed algorithm is based on the extraction of the pupil’s location from an image~\cite{19_gupta}. When the pupil is located, Cartesian plane coordinates are used to track the pupil.

Eye tracking to control a cursor involves pupil detection, and thereafter establishing direction of cursor movement. The algorithms and methods used for these processes are discussed in the pupil detection, direction determination, and eye position tracking subsections.

\subsection{Pupil Detection}

Pupil detection was chosen instead of iris detection for two reasons: standard pupil shape, and variation in iris colour. The pupil is ordinarily seen as a full circle, whereas the top of the iris is often covered by the upper eyelid. This makes detection difficult because of shape variance. The iris can be a variety of colours, which complicates edge detection as the image must be converted to a binary image. The threshold for generating the binary image requires user-specific calibration – for example, the threshold is far lower with blue eyes than dark brown eyes. The pupil, however, is a constant for every user’s eye. Hence, using pupil detection greatly simplifies the system.
The camera produces a greyscale image due to the optical light filter and infrared light source. To complete the edge detection, the original greyscale image is converted to a binary image. The appropriate threshold is determined by noting the light intensity in the room, which can be auto-calibrated. The thresholding process results in a binary image containing a black circle. Depending on light intensity changes in the room, the binary image may contain artefacts that appear as black spots.

The algorithm used to detect pupil edges is the OpenCV findContours() function~\cite{25_Suzuki_Topological_1985}. This function locates contours in a binary image. Each contour is a vector of points that is stored in a vector of contours. The final steps in pupil detection rely on two pupil characteristics: colour (black) and shape (circular).

Each contour in the vector of contours is analysed. First, to determine if the contour is large enough to be the pupil. This prevents the system from detecting small black flecks in the image that come from noise and changes in light intensity (artefacts) . Second, to establish if the contour is circular. Further work is required for elliptical contour detection. Circular contour detection is performed in two steps: creating a boundary around the contour, and calculating the area of the contour. A bounding rectangle is created around the contour. If the contour is circular, the width and height of the bounding rectangle should be the same, indicating that the boundary is a square. Therefore, dividing height by width results in unity. The area of the contour is found using the OpenCV contourArea() function. For pupil detection, under the condition that the pupil is circular, the area found should reflect that determined when Equation 1 is used.

\begin{equation} \label{eq:1}
area = \pi\times r^2
\vspace{3.5mm}
\end{equation}

Dividing the expected area by the area found using \textit{contourArea()} results in unity if the contour is a circle. However, the results cannot be compared to unity because the bounding rectangle covers a larger area than the circle. Therefore, each result is subtracted from unity, and com-pared to 0.2 (approximate heuristic value). This accounts for any errors caused by the bounding rectangle and \textit{contourArea()}. Once the pupil is detected, the OpenCV \textit{drawContours()} function is used to draw a circle around the pupil. The system determines the direction in which the eye is moving.

\subsection{Direction Determination}

In the incremental tracking of smooth eye movements, the head and eyes move in opposite directions so that the gaze remains fixed on a single point. This is used to move the cursor in the desired direction using incremental tracking of smooth eye movements.

When the system is initialised, the mouse cursor moves to the centre of the screen. The user must look at the cursor so that the position of the pupil is captured for calibration purposes. The central position of the mouse cursor provides a baseline image that is stored. The pupil is detected in the baseline image and the position of the pupil is stored. Once calibrated, the original pupil position can be compared to the newly detected pupil position to find the direction that it moves.

Subsequently, the direction of the pupil must be found as the head moves. The new position of the pupil, x and y coordinates, are subtracted from the corresponding x and y coordinates of the baseline pupil position, as shown in Equation 2. The resulting coordinates, which will be referred to as the difference coordinates, are compared to the baseline coordinates. If the difference x coordinate, $x_{difference}$, is less than the baseline x coordinate, $x_{baseline}$, the new x position, $x_{newPosition}$, is the result of adding $x_{baseline}$ and $x_{difference}$, as in Equation 3. The y coordinates are calculated using the same procedure.

\begin{equation} \label{eq:2}
x_{difference} = x_{baseline} - x_{newPosition}
\end{equation}

\begin{equation} \label{eq:3}
x_{newPosition} = x_{difference} + x_{baseline} \vspace{3.5mm}
\end{equation}

All conditions and corresponding calculations necessary to determine the new cursor position are reflected in Table~\ref{tab:cursor_pos}. This sets the new position for the cursor.

\begin{table}[!t]
\renewcommand{\arraystretch}{1.3}
\caption{New cursor positions}
\label{tab:cursor_pos}
\centering
\begin{tabular}{|c||c|}
\hline
Condition & New position\\
\hline
$x_{difference} < x_{baseline}$ & $x_{difference} + x_{baseline}$\\
\hline
$x_{difference} > x_{baseline}$ & $x_{difference} - x_{baseline}$\\
\hline
$y_{difference} < y_{baseline}$ & $y_{difference} - y_{baseline}$\\
\hline
$y_{difference} > y_{baseline}$ & $y_{difference} + y_{baseline}$\\
\hline
\end{tabular}
\end{table}

For example, if the baseline pupil position is (25, 25) and the new pupil position is found to be (35, 25), $x_{difference}$, calculated using Equation~\ref{eq:2}, is -10, and $y_{difference}$ is 0.

Table~\ref{tab:cursor_pos} presents the conditions under which the new x and y coordinates are determined. In the example, both the x and y difference coordinates are less than the corresponding baseline coordinates, indicating that the new position for the cursor, determined using the new position formulae provided in Table~\ref{tab:cursor_pos}, is (15,-25). All other coordinates are found using the same procedure, with the new position coordinates determined according to conditions presented in Table~\ref{tab:cursor_pos}.


\subsection{Eye Position Tracking}
\textit{Pupil}, an open source gaze tracking software described by~\cite{Kassner}, was used to compare the smooth eye tracking system with saccadic eye tracking. Pupil was chosen because of the simplicity with which it integrates with the custom built PlayStation-Eye camera. It has a custom-built calibration setting to track the pupil in order to visualise the gaze positions with a cross (x). This was then used as a reference point to move the mouse cursor using the position determined by eye tracking.
When the application is first run, the user looks directly into the camera for calibration purposes. Thereafter, the user has full freedom of head movement. The pupil is assumed to be fully visible; it is always a full circle, never covered by the upper eyelid. Further algorithmic development and investigation is required for situations where the pupil is partially obscured, but that is beyond the scope of this work.

\section{Testing and results}

\subsection{Experimental Procedure}

The two participants in the study were two of the authors (male and female), who sat upright on a chair reclined at a $100~^{\circ}$ angle~\footnote{This version of the manuscript has been updated to reflect latest data}. The chair was adjusted to ensure that both feet were flat on the ground. The computer monitor was positioned directly in front of the participant’s eyes, at eye level, at a distance of 50 cm. The participants then used the system to test for certain system characteristics: accuracy; precision; and time to track a position.

The study was designed to optimise the quality of the pupil-position data. The constructed headgear with the PlayStation-Eye camera, as in Figure 1, was worn by each participant. The infrared light was shone on the participant’s left eye and captured by the PlayStation-Eye camera.

The system was initially tested for accuracy and time characteristics. Measurement collection was repeated and compared to those of Pupil. Each test was repeated ten times, and the results were averaged.

\subsection{Accuracy}

To test for accuracy, four squares were drawn with the following dimensions in pixels: $100\times100$; $50\times50$; $25\times25$; and $9\times9$. The aim was to move the cursor into the centre of each square. The squares were placed in the centre of the screen with coordinates (960, 540). The user had to move the cursor from the top-left corner, coordinates (0,0), to within the square. This distance is approximately 30.4 cm, and the cursor must move at an angle of $135~^{\circ}$ relative to the top-left corner of the monitor. The process was repeated for improved accuracy of results. Both the smooth tracking and saccadic tracking systems were measured for comparison.

Results are displayed in Table 2, indicating the accuracy. Accuracy is defined as the percentage of times the user was able to place the cursor inside the boundaries of each square. Smooth eye movement tracking consistently produced 100 \% accuracy of results for the larger three square dimensions, and 90~\% for the $9\times9$ square. In contrast, saccadic tracking’s accuracy decreased as the square dimensions decreased, producing $75~\%$, $65~\%$, and $5~\%$ accuracy for the $100\times100$, $50\times50$, and $25\times25$ squares, respectively, and $0~\%$ accuracy for both the $9\times9$ squares.

\begin{table}[!t]
\renewcommand{\arraystretch}{1.3}
\caption{Accuracy of test results}
\label{tab:accuracy}
\centering
\begin{tabular}{|c|c|c|}
\hline
Square size (pixels) & Smooth eye
tracking (\%) & Saccadic eye tracking (\%)\\
\hline
$100\times100$ & 100 & 75 \\
$50\times50$ & 100 & 65 \\
$25\times25$ & 100 & 5 \\
$9\times9$ & 90 &   0 \\
\hline
\end{tabular}
\end{table}

\subsection{Resolution}

The average distances from the centres of the squares when using the smooth tracking and saccadic tracking systems were 27.9 and 140.1 pixels, respectively. Equation 4 is used to convert the measurements from pixels to centimetres, resulting in a ratio of $0.0276~cm/pixel$.

\begin{equation} \label{eq:ratioofcmtopixels}
\textrm{cm:pixel ratio} = \frac{\textrm{screen length in cm}}{\textrm{screen length in pixels}} \vspace{3.5mm}
\end{equation}

The average distance when using the smooth eye tracking system was 0.80~cm away from the centre of the square, whereas, when using saccadic tracking, the average distance was 3.22~cm – a substantial difference between the two tracking methods. The average distance of the smooth eye tracking system relative to that of the saccadic tracking system shows a 400~\% increase.

Precision can partially be inferred given that each square size measurement was tested ten times per user.

\subsection{Time to Track Position}

\begin{figure}[!t]
\centering
\includegraphics[width=4in]{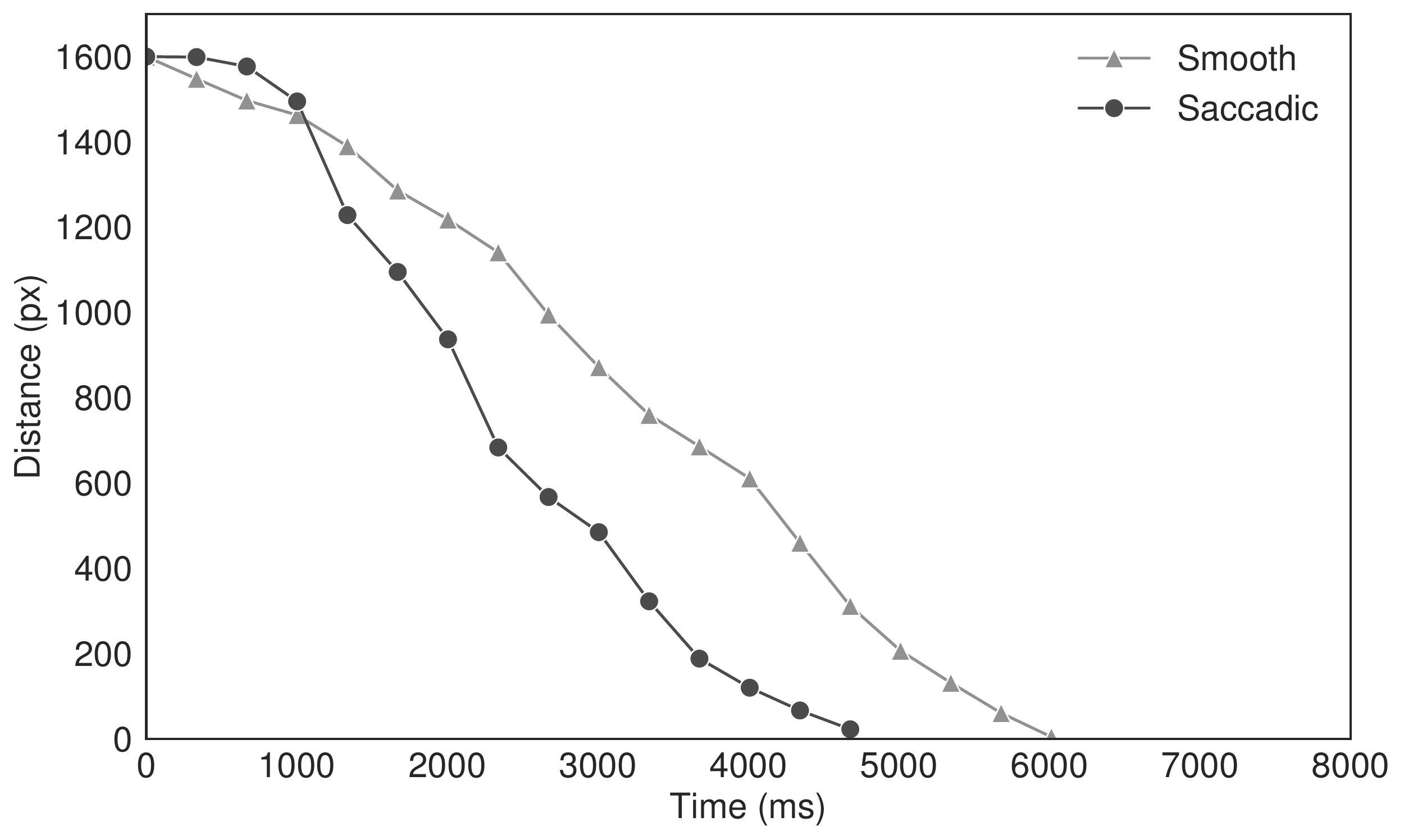}
\caption{Graph illustrating time taken for the cursor to move when using both incremental and gaze tracking systems}
\label{fig:trackingvstime}
\end{figure}

The time taken for a user to move a mouse from one position to another is measured. For each test, the cursor started at an identical position and the user moved the cursor to a designated end point. The average time taken when using the smooth eye tracking system was 6.45~s, and 2.61~s when using the saccadic tracking system. This is illustrated in Figure~\ref{fig:trackingvstime}.

Time tests were implemented in the software. The timer was initiated when the pupil first moved, and stopped when the cursor entered a specific region of pixels. This ensured more accurate results than timing by hand, where reaction time must be considered. Figure 4 shows the tracking curve for both the smooth eye track-ing and the saccadic tracking systems. The tracking curve for the smooth eye tracking system shows a more direct line from the start position to the end position. This indicates that the user had greater control over the cursor’s position and the direction in which it moved with the closed-loop, saccade-free, smoother motion. The saccadic tracking curve shows the cursor moving in a straight line with a positive gradient. The cursor then moved vertically downwards to the end position. The saccadic tracking algorithm is generally open-loop and faster, which com-promises control and hinders the user’s ability to rapidly correct the cursor’s trajectory.

\begin{figure}[!t]
\centering
\includegraphics[width=4in]{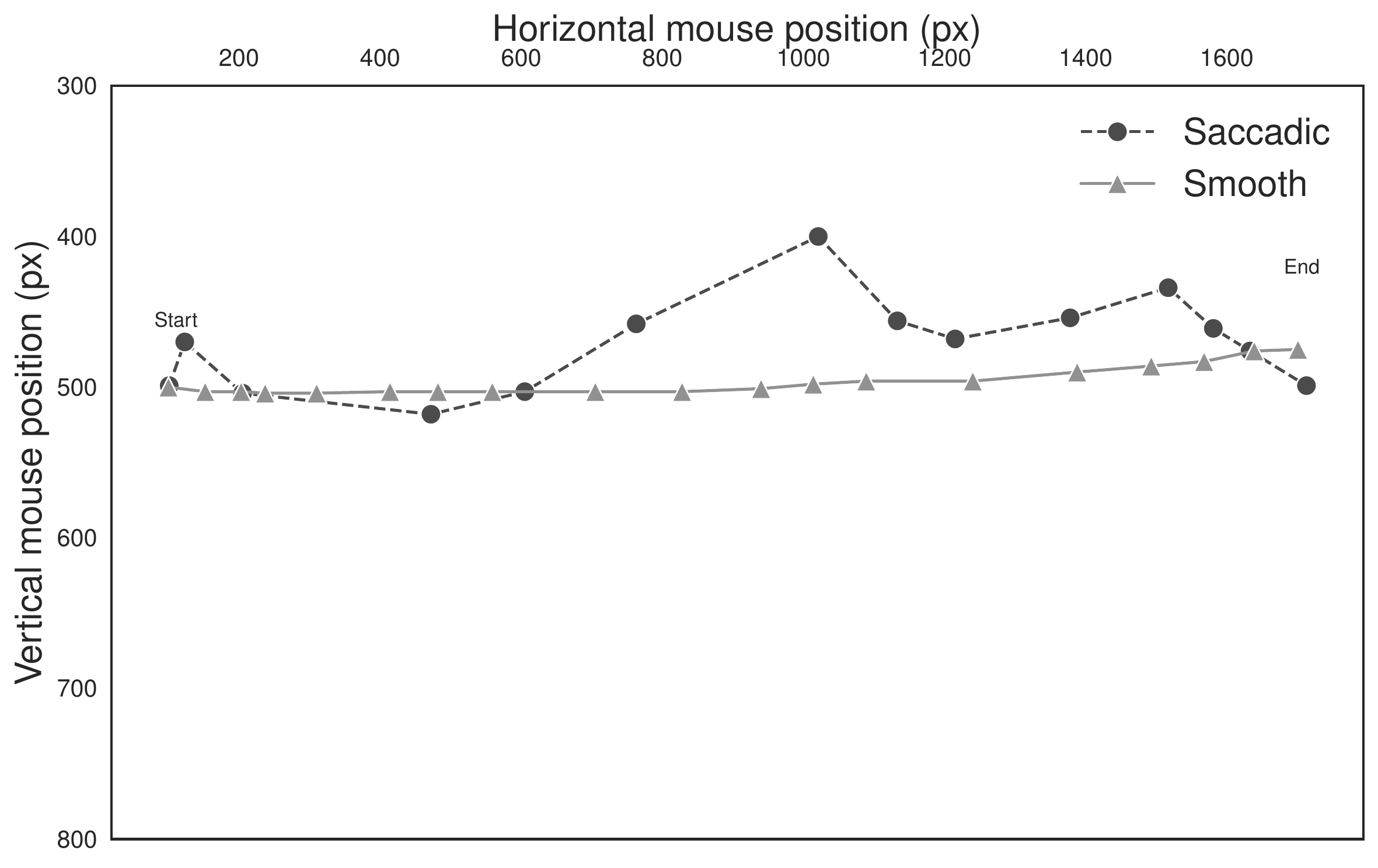}
\caption{Graph depicting the tracking curve of the cursor for both smooth eye tracking and saccadic tracking systems.}
\label{fig:mousecurve}
\end{figure}

\subsection{Pupil Detection Range}

The pupil is not always perceived as a circle. When it is at the extreme lateral or medial margins of the orbit, the pupil appears to be elliptical in shape. This affects the pupil detection range. The range was calculated to be approximately 12~\degree. The user, however, remained able to position the cursor everywhere on the screen. Further work is required to improve pupil detection, but that is outside the scope of this primarily proof-of-concept study.

\subsection{Smooth Tracking and Saccadic Tracking System Characteristics}

The movement of the mouse cursor using smooth eye tracking and saccadic tracking have the same characteristics because both systems are based on the same hardware and software algorithms. The lens of the PlayStation-Eye camera has a $23~\degree$ field of view which translates to $0.012~\degree/px$ or $83~px/\degree$ on a $1920\times1080$ screen.


\section{Discussion}

The system presented exploits the VOR to provide a novel solution to the inaccuracies in saccadic tracking systems, evident in Figures~\ref{fig:trackingvstime} and~\ref{fig:mousecurve}. The VOR-related smooth eye tracking system is approximately~four times more accurate than the saccadic tracking system. Precision tests indicate that the saccadic tracking system had a 75~\% success rate for a $100\times100$ pixel block, whereas the smooth eye tracking system had a $90~\%$ success rate for a $9\times9$ pixel block.

The smooth eye tracking system creates a better experience for the user because it offers finer, more controlled handling of the cursor. The improved handling is partly due to the inherently slow smooth eye movement. By forcing a slower speed of eye movement, users have enhanced abilities to: drag the cursor using head movements; continuously monitor the cursor on its trajectory; and control cursor position with greater ease. This allows the user to manipulate the cursor as frequently as required, and the user is not hindered by the lack of control. The smooth eye tracking system is simpler than the saccadic tracking system because it does not have to compensate for the inaccuracies inherent in saccadic tracking.

The time test revealed that the smooth eye tracking system allowed the cursor to reach its end point in approximately $40~\%$ of the speed of saccadic tracking. This result is expected because of the respective speeds of different eye movements used. Slower speed increases user control over the cursor. Additionally, the smooth eye tracking movement provides the user with better visual feedback because, as the cursor moves, the eyes move with it. With the saccadic tracking system, however, the user must look away from the cursor. This causes a time delay for the visual feedback and further feedback due to saccades.

The headgear is cumbersome and interferes with the user’s view of the screen, despite efforts to prevent this. The solution presented cannot distinguish between intentional smooth eye tracking movements and general movements of the eye, which cause the cursor to move slightly with any slight eye movement. The algorithm cannot detect the pupil when it is at the extreme lateral or medial margins of the orbit because it no longer appears circular, but rather is elliptical. This diminishes the rate of cursor movement, but does not detract from the range of movement.

Applications of such a system include facilitating computer use by people who are otherwise unable to manipulate a hand-controlled mouse. With the addition of clicking capabilities, the system may allow users to interact more easily with a computer, and it can create better gaming experiences in first-person games – the player can alter their point-of-view by changing the direction of their head. The system was tested while playing Counter Strike, and the user was able to move the player’s visual point-of-view naturally and easily, creating an immersive, enjoyable gaming experience.  Other applications include: remote control for industrial equipment; surgical applications; controlling Head-up Displays; and possibly in defence.

Saccadic tracking is cheaper as it can be implemented using a web camera, which ordinarily comes with standard computers. Additionally, it is a more intuitive way of moving the cursor, whereas using smooth eye tracking initially requires time to adjust to the procedure. Mouse inversion, however, may resolve this issue, but that is not tested in this work.

Infrared rather than visible light was chosen to simplify image processing, however, this required additional safety considerations.


\subsection{Recommendations for Future Work}
The system can be improved by increasing the range in which a pupil can be detected, specifically by introducing elliptical pupil detection. The headgear on which the camera is mounted can be improved or redesigned to increase user comfort. The thresholding constant can be dependent on light intensity so that pupil detection remains constant. A function can be introduced to distinguish between smooth (incremental) eye movements and general eye movements for when the user is reading text.

\section{Conclusion}

A system that allows movement of a cursor with the use of VOR-related smooth eye movement using incre-mental tracking was successfully developed and compared to saccadic, step change, eye tracking. Hardware consists of headgear with a PlayStation-Eye camera that has been modified to perceive exclusively infrared light, an infra-red light source, and a computer. The software subsystem allows the images received from the camera to be analysed. The analysis includes pupil detection, finding the direction of pupil movement, and moving the cursor the appropriate distance and direction. Testing indicated that, when compared to saccadic eye tracking systems, the smooth eye tracking system has greater accuracy and precision. However, due to the slow, incremental nature of smooth tracking, the cursor moves slower than when saccadic eye tracking is used.

\bibliography{bare_jrnl}

\end{document}